\documentclass[a4paper,fleqn,usenatbib]{mnras}

\usepackage{newtxtext,newtxmath}

\usepackage[T1]{fontenc}
\usepackage{ae,aecompl}
\usepackage{relsize} 
\usepackage{color,soul}
\usepackage{subfig}

\usepackage{graphicx}
\usepackage{amsmath}
\usepackage{amssymb}
\usepackage{bm}
\usepackage{footnote}
\usepackage{bigints}
\usepackage{etoolbox}
\makeatletter
\AtBeginEnvironment{thebibliography}{%
   \clubpenalty10000
   \@clubpenalty \clubpenalty
   \widowpenalty10000}
\makeatother
\defcitealias{Eis18}{E18}

\title[External photoevaporation of PPDs in $\sigma$ Orionis]{\vspace{-4mm}Testing viscous disc theory using the balance between stellar accretion and external photoevaporation of protoplanetary discs\vspace{-3mm}}

\author[Winter et al.]{Andrew J. Winter$^{1,2}$\thanks{andrew.winter@uni-heidelberg.de}, Megan Ansdell$^{3,4}$, Thomas J. Haworth$^5$,\newauthor and J.~M.~Diederik~Kruijssen$^{1}$\\
$^{1}$Astronomisches Rechen-Institut, Zentrum f\"{u}r Astronomie der Universit\"{a}t Heidelberg, M\"{o}nchhofstra\ss e 12-14, 69120 Heidelberg, Germany	\\
$^{2}$School of Physics and Astronomy, University of Leicester, Leicester, LE1 7RH, UK\\
$^3$Flatiron Institute, Simons Foundation, 162 Fifth Ave, New York, NY 10010, USA \\
$^4$NASA Headquarters, 300 E Street SW, Washington, DC 20546, USA \\
$^5$Astronomy Unit, School of Physics and Astronomy, Queen Mary University of London, Mile End Road, London, United Kingdom \\
\vspace{-7mm}
}

\date{Accepted 2020 {June} 5. Received  2020 {May} 5; in original form 2020 May 27\vspace{-2mm}}

\pubyear{2020}

\begin{document}
\label{firstpage}
\pagerange{\pageref{firstpage}--\pageref{lastpage}}
\maketitle

\begin{abstract}
The nature and rate of (viscous) angular momentum transport in protoplanetary discs (PPDs) has important consequences for the formation process of planetary systems. While accretion rates onto the central star yield constraints on such transport in the inner regions of a PPD, empirical constraints on {viscous spreading} in the outer regions remain challenging to obtain. Here we demonstrate a novel method to probe the angular momentum transport at the outer edge of the disc. This method {applies to} PPDs that {have lost a significant fraction of their mass due} to thermal winds driven by UV irradiation from a neighbouring OB star. We demonstrate that this external photoevaporation can explain the observed depletion of discs in the $3{-}5$~Myr old $\sigma$ Orionis region, and use our model to make predictions motivating future empirical investigations of disc winds. For populations of intermediate-age PPDs, in viscous models we show that the mass flux outwards due to angular momentum redistribution is balanced by the mass-loss in the photoevaporative wind. A comparison between wind mass-loss and stellar accretion rates therefore offers an independent constraint on viscous models in the outer regions of PPDs.

\end{abstract}

\begin{keywords} 
stars: formation, circumstellar matter -- protoplanetary discs\vspace{-3mm}
\end{keywords}


\section{Introduction}

Planets form and spend their early phases of evolution in protoplanetary discs (PPDs) composed of dust and gas. The dispersal of the gaseous component occurs over a time-scale of  $\sim 5$--$10$~Myr \citep[e.g.][although aging young stars is infamously challenging; see for example \citealt{Bel13}]{Hai01}, during which material is depleted both by accretion onto the central star \citep[e.g.][]{Man16, Man17}, and thermally or magnetically driven winds \citep[e.g.][]{Erc17}. Prior to PPD dispersal, several aspects of planet formation, growth and evolution remain poorly understood. One important example is angular momentum transport through the disc, which (among others) affects dust growth and radial drift \citep[e.g.][]{Bir12} and planet migration \citep[e.g.][]{Ale09}. Despite its importance, it remains unclear whether angular momentum is radially redistributed (i.e. viscous transport) or removed \citep[e.g. by magnetohydrodynamic winds --][]{Blandford82, Bai13}. Molecular viscosity is far too low to yield observed accretion rates, and alternative mechanisms for viscous transport include magnetorotational \citep[][]{Bal91, Les14} or gravitational torques \citep[][]{Clarke09, Rice10} and back-reaction of solids on the gaseous disc \citep[e.g.][]{Dip18}. It remains unclear if models such as the frequently applied $\alpha$-viscosity models \citep{Sha73}, are appropriate for describing disc evolution. While it is possible to measure stellar accretion rates, similar constraints for mass redistribution in the outer regions of PPDs are more challenging. An ALMA survey by \citet{Taz17} revealed a greater radial extent of dust discs than in Lupus than in younger star-forming regions such as Taurus-Auriga and Ophiuchus, suggesting viscous expansion. However, \citet{Bar17} find comparatively compact discs in the older Upper Sco region, leading to the opposite conclusion. Any evidence from dust disc radii must be considered with caution, both because the dust may not be physically coupled to the gas and because opacity effects can lead to  differences between observed and physical radii \citep{Ros19}. While gas disc radii may be a more promising diagnostic \citep{Tra20}, the number of resolved CO detections remains small, particularly when controlling for ages of star forming regions. In this Letter, we explore a more direct method, which does not depend on inferring disc outer radii, to measure angular momentum transport in the outer regions of PPDs.

Observational and theoretical evidence supports the influence of external photoevaporation due to extreme- and far-ultraviolet (EUV/FUV) irradiation of PPDs with a neighbouring OB star 
(e.g. \citealt{Ada04}; \citealt{JOv12}; \citealt{Fan12}; \citealt{Fac16}; \citealt{Gua16}; \mbox{\citealt{Win18b}}; \citealt{Nic19}; \citealt{Con19}). Early investigations were restricted to the bright proplyds in the Orion Nebula Cluster \citep[ONC;][]{Ode94}, but recent studies have uncovered examples of proplyds \citep{Kim16} and dust mass depletion \citep{Ans17} in regions experiencing much lower UV fluxes. Theoretical mass-loss rates \citep[e.g.][]{Haw18b} suggest more than half of discs in the solar neighbourhood are significantly influenced by external photoevaporation \citep{Win20}. However, the highest mass-loss rates of $10^{-6} \, M_\odot$~yr$^{-1}$ observed in the brightest ONC proplyds probably represent short-lived periods of extreme disc depletion \citep{Win19b}. Extended discs are rapidly truncated during such periods; \citet{Ros17} showed how, for a photoevaporated disc population of a certain age, this truncation affects the inferred viscous time-scale (the ratio of the disc mass $M_\mathrm{disc}$ to accretion rate $\dot{M}_\mathrm{acc}$). 

In this Letter, we demonstrate how a disc population that has been depleted by external photoevaporation can be used to probe angular momentum transport in the outer disc by comparing the externally driven mass-loss rate in the wind $\dot{M}_\mathrm{wind}$ with $\dot{M}_\mathrm{acc}$. If the outer part of the disc has already been sufficiently eroded, then $\dot{M}_\mathrm{wind}$ is regulated by the rate at which mass can be replenished in the outer part of the disc; this is dependent on the rate of angular momentum transport. To demonstrate this principle, we choose the $\sim 3{-}5$~Myr old $\sigma$ Orionis star forming region \citep[][although an older or younger age is possible -- \citealt{Bel13,Kou18}]{Oli02, Oli06} as an illustrative example, motivated by recent evidence of environmentally depleted PPDs \citep{Ans17} and theoretical studies suggesting wind tracers at the moderate FUV fluxes characteristic of the region \citep{Haw20}. We show that if there were to be future observational evidence for mass-loss in a
photoevaporative wind at a rate comparable with the accretion rate onto the star then it
would imply PPDs undergo viscous diffusion.


\vspace{-0.2cm}
\section{Models}
\label{sec:model}
We calculate the surface density of PPDs undergoing both viscous evolution and depletion by external photoevaporation by tracking the UV flux experienced by each disc over the course of an $N$-body integration. Our methods are detailed in \citet{Win19b}, and we briefly summarise them here.

\subsection{Stellar dynamical model}

We calculate the dynamical evolution of a stellar population representative of that in $\sigma$ Orionis using the \textsc{Nbody6++} code \citep{Aar74}. We draw initial star particle positions $r$ from a distribution according to a \citeauthor{Plu11} density profile:
\begin{equation}
\label{eq:dprof}
    \rho_*(r)  = \rho_0 \left( 1 + \frac{r^2}{a^2}\right)^{-5/2},
\end{equation}where we choose $a=0.32$~pc to yield a half-mass radius of $0.4$~pc consistent with the observed density profile \citep{Cab08}. We fix the normalisation constant $\rho_0$ such that the total stellar mass is $150 \, M_\odot$, with a \citet{Kro01} initial mass function (IMF):
\begin{equation}
\label{eq:imf}
\xi(m_*) \propto \begin{cases}
               m_*^{-1.3} \quad \mathrm{for } \, 0.08 \, M_\odot \leq m_*< 0.5 \, M_\odot\\ 
               m_*^{-2.3} \quad \mathrm{for } \, 0.5 \, M_\odot \leq m_* < 12 M_\odot \\
              0 \qquad \quad \, \mathrm{otherwise}
            \end{cases}.
\end{equation}The upper mass limit of $12 \, M_\odot$ is chosen such that FUV flux is dominated by the most massive star in the region, the $17 \, M_\odot$ component of the $\sigma$ Ori multiple system \citep{Sch16}. We place a $17 \, M_\odot$ star at the centre of the region, and define an isotropic Maxwellian velocity distribution, with a dispersion chosen such that the entire stellar population is in virial equilibrium. 

\subsection{Disc evolution}

To calculate the disc evolution we follow \citet{Cla07} in solving the viscous diffusion equation across a grid uniformly spaced in $R^{1/2}$, where $R$ is the radial coordinate of the disc, with a viscosity $\nu \propto R$, equivalent to a constant \citeauthor{Sha73} $\alpha$ \citep[see also][]{Hartmann98}. We adopt the same disc initial conditions as applied in \citet{Win19b} to reproduce the properties of PPDs in the Orion Nebula Cluster, such that models can be compared (Section~\ref{sec:mdotwindacc}). The initial surface density profile is defined with a scale radius $R_\mathrm{s}=20$~au, and truncated outside of $R=50$~au. We choose a viscous $\alpha=10^{-3}$ (see \citealt{Win19b} and Section~\ref{sec:caveats} for a discussion on the influence of varying $\alpha$), and assume the initial disc mass $M_\mathrm{disc} = 0.1 \, m_*$. We calculate the evolution of individual PPDs until the mass $M_\mathrm{disc}<10^{-5} \, M_\odot$, below which the disc is considered to be dispersed. 

Mass-loss induced by the photoevaporative wind occurs at the disc outer edge ($R_\mathrm{disc}$) at a rate:
\begin{equation}
    \dot{M}_\mathrm{wind} = \max \left\{\dot{M}_\mathrm{EUV}, \dot{M}_\mathrm{FUV} \right\} 
\end{equation}where $\dot{M}_\mathrm{EUV}$ and $\dot{M}_\mathrm{FUV}$ are the EUV and FUV induced mass-loss rates respectively. \citet{Joh98} analytically express the EUV mass-loss rate:
\begin{equation}
\label{eq:EUVloss}
\frac{\dot{M}_\mathrm{EUV} }{ M_\odot \, \mathrm{yr}^{-1}} \approx 6.3 \times 10^{-8} \left( \frac{d}{0.1\, \mathrm{pc}}\right)^{-1} \left(\frac{ \Phi_\mathrm{i}}{10^{49} \, \mathrm{s}^{-1}} \right)^{1/2}  \left( \frac{R_\mathrm{disc}}{50 \, \mathrm{au}}\right)^{3/2},
\end{equation} while $\dot{M}_\mathrm{FUV}$ is obtained by interpolating over the \textsc{FRIED} grid \citep{Haw18b}. The \textsc{FRIED} grid has a floor at $10^{-10}\, M_\odot$~yr$^{-1}$, and we therefore assume free expansion of the disc (no mass-loss) if this floor is reached. FUV photons statistically determine $\dot{M}_\mathrm{wind}$ for the majority of winds from PPDs, while EUV induced mass-loss may cause initial rapid depletion when the ionisation front is close to the disc edge \citep[i.e. the limit of a thin photodissociation region;][]{Joh98}.

The flux experienced by the disc around each star is calculated by tracking the distance to every other star with mass $m_*>1\, M_\odot$. The luminosity and effective temperatures of all such stars are taken from \citet{Sch92} for metallicity $Z = 0.02$ and at the time closest to $1$~Myr (although our results are not sensitive to this choice). Atmosphere models by \citet{Cas04} give the wavelength-dependent luminosity, which we then integrate  over FUV ($6$~eV~$< h\nu < 13.6$~eV) and EUV ($h \nu> 13.6$~eV) energy ranges. We do not include a prescription for interstellar extinction.
\begin{figure}
    \centering
    \includegraphics[width=\columnwidth]{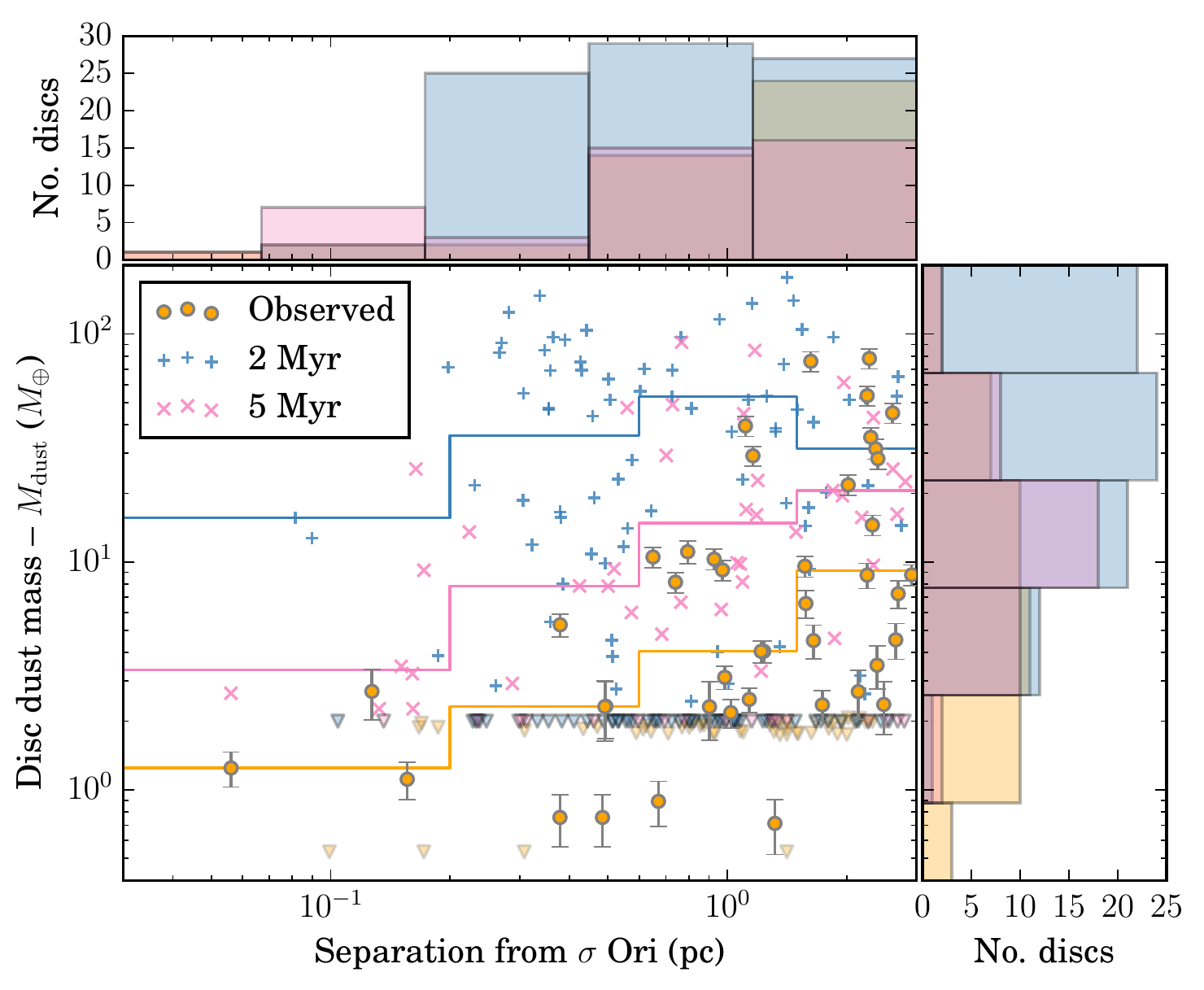}
    \vspace{-0.6cm}    
    \caption{Protoplanetary disc dust masses inferred from $1.3$~mm flux  \citep[orange points;][]{Ans17, Ans20} and the disc masses obtained from the dynamic model with an assumed dust-to-gas ratio of $10^{-2}$ at $2$~Myr (blue `+' crosses) and $5$ Myr (pink `x' crosses) versus projected separation from the most massive star in the region. For the observational data, upper limits (non-detections) are shown as faint downward triangles symbols, while detections are shown as circles with error bars. We treat all surviving discs with dust masses $<2\, M_\oplus$ from our model as `non-detections' (upper limits; faint downward triangles) for direct comparison, with the equivalent colours as the `detections'. All bars are for detected discs only, with those in the scatter plot representing the median of the masses in each distance bin. The top panel shows the distribution of detections with distance from $\sigma$ Ori and the right panel shows the distribution of dust mass detections.}
    \vspace{-0.4cm}    
    \label{fig:sepvmdust}
\end{figure}

\vspace{-0.2cm}
\section{Results and Discussion}

In Section \ref{sec:discmass} we demonstrate how the model we have applied successfully reproduces the distribution of PPD masses inferred from $1.3$~mm {ALMA} surveys of discs in $\sigma$ Orionis by \citet{Ans17, Ans20}. In Section~\ref{sec:mdot} we make predictions for photoevaporative mass-loss rates in the region to illustrate how observations may be used to constrain disc physics. 

\subsection{Disc masses}
\label{sec:discmass}
\begin{figure*}
    \centering
    \includegraphics[width=0.84\textwidth]{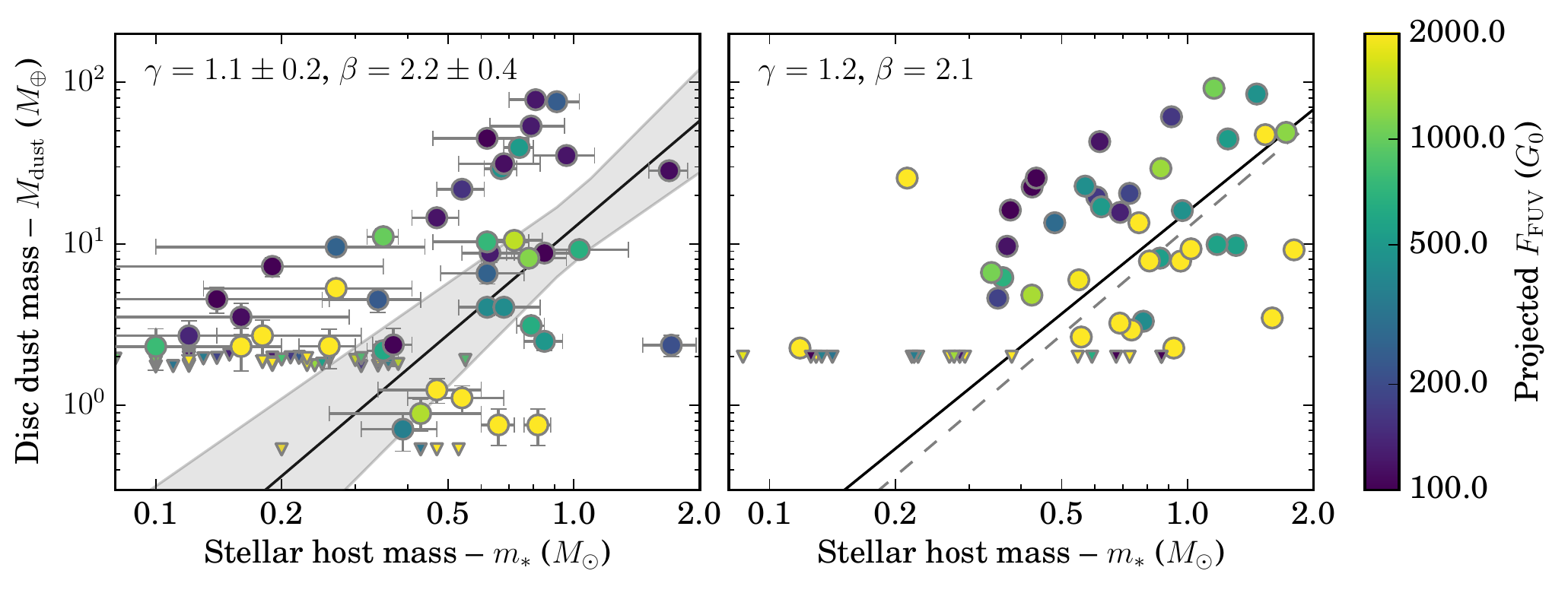}
    \vspace{-0.3cm}    
    \caption{Power law relationship fits between disc dust masses and stellar host mass, $\log M_\mathrm{disc}/M_\oplus = \gamma + \beta \log m_*/M_\odot$. The left panel is for observed disc hosting stars, while the right is for the model after $5$~Myr with an assumed dust-to-gas ratio of $10^{-2}$. In both cases, points are colour coded by the FUV flux that is estimated from the projected distance from the most massive ($m_* = 17\, M_\odot$) star. The observed relationship \citep{Ans20} is obtained from the \textsc{Linmix} fitting procedure \citep[left panel solid line, right panel faint dashed line;][]{Kel07}, and we repeat this procedure to obtain the power law relationship shown by the solid black line in the right panel for our model. Dust masses $<2\, M_\oplus$ in the model are treated as upper limits for fitting purposes.}
         \vspace{-0.4cm} 
    \label{fig:mstvmdust}
\end{figure*}

  \citet{Ans17} found the dust mass of PPDs decreases with proximity to $\sigma$ Ori, strongly suggesting that the discs have been depleted by UV irradiation. In Figure~\ref{fig:sepvmdust}, we compare the observed dust masses with those obtained from our model (scaling by a dust-to-gas ratio of $10^{-2}$) at $2$~Myr and $5$~Myr. After $2$~Myr, the discs in the model have not yet been sufficiently depleted to demonstrate a clear gradient in disc masses, while after $5$~Myr the masses show signatures of depletion close to $\sigma$ Ori, comparable to the observed mass distribution. At $5$~Myr, disc masses in the model are consistently greater than observed masses by a factor $\sim 2$ at each distance bin, which could be a due to dust processing, internal depletion or observational effects \citep[see Section~\ref{sec:caveats} and][]{Taz17}.

  The power-law relationship between disc and stellar host is
    \begin{equation}
        \label{eq:mdiscmhost}
       \log \left(\frac{ M_\mathrm{disc}}{1\, M_\oplus}\right) = \gamma +\beta  \log \left(\frac{m_*}{1\, M_\odot} \right),
    \end{equation}
    where the index $\beta$ increases with stellar age \citep{Bar16, Pas16,Ans16,Ans17}. The steepening of this relationship may be driven by either internal processes {\citep[e.g. dust growth and evolution;][]{Pas16}} or by external photoevaporation, since circumstellar material is thermally unbound more efficiently for a shallower gravitational potential. In either case, we aim to approximately reproduce the relationship to make predictions for the expected mass-loss rate distribution in Section~\ref{sec:mdot}. We use the \textsc{Linmix} package to fit equation~\ref{eq:mdiscmhost} to our model results after $5$~Myr of evolution and show the result in Figure~\ref{fig:mstvmdust}. The equivalent dust masses and stellar mass, with power-law fit are taken from \citet{Ans20}. We draw uncertainties in $M_\mathrm{disc}$ and $m_*$ in our model from the observational uncertainties, and treat all discs with a dust mass $< 2 \, M_\oplus$ as upper limits. We obtain $\gamma = 1.2$ and $\beta =2.1$, consistent with $\gamma =1.1\pm 0.2$ and $\beta = 2.2\pm 0.4$ for the observed dust mass estimates \citep{Ans20}.
    
We conclude that our model satisfies observational constraints on the disc mass distribution in $\sigma$ Orionis. Hence, external photoevaporation is a viable mechanism for disc depletion in the region. Our model also predicts the expected photoevaporative mass-loss rates $\dot{M}_\mathrm{wind}$ as a function of stellar and disc properties.

\begin{figure}
    \centering
    \includegraphics[width=0.95\columnwidth]{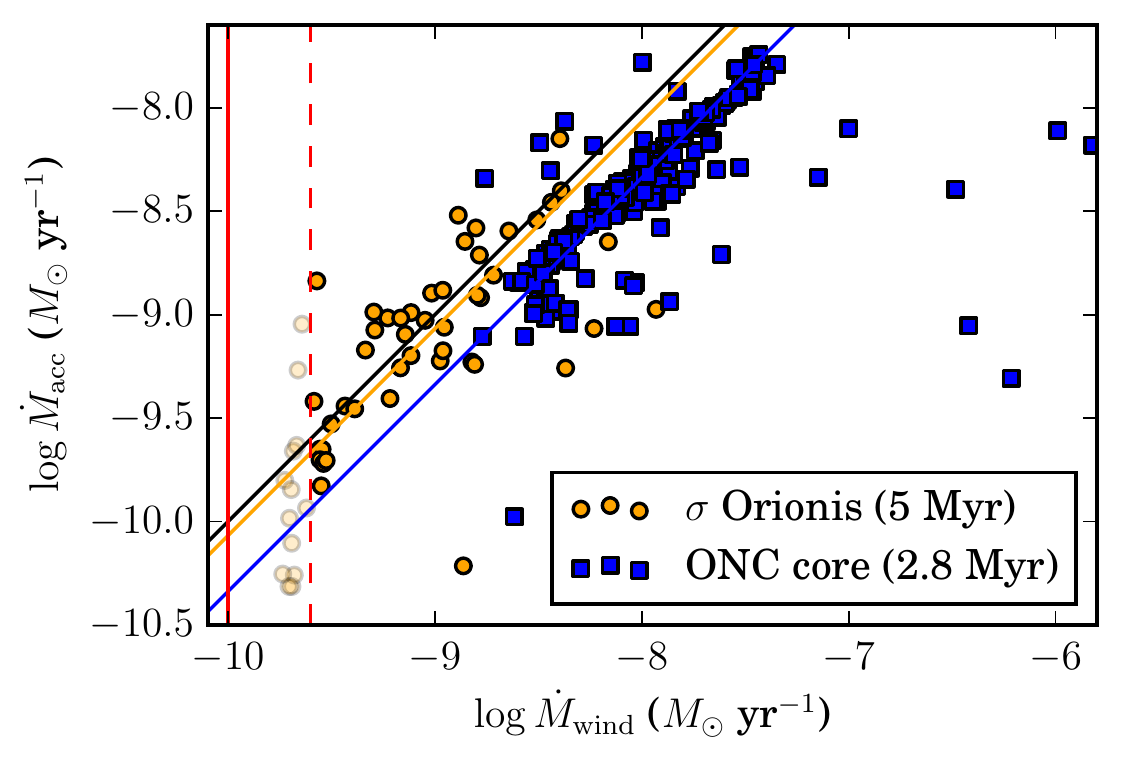}
    \vspace{-0.45cm}
    \caption{{Accretion rate versus wind-induced mass-loss rate obtained from models for PPDs in $\sigma$ Orionis and the core ($<0.3$~pc of $\theta^1$C) of the ONC \citep[][]{Win19b}. Excluding the points to the left of the dashed line ($\dot{M}_\mathrm{wind}< 2.5\times 10^{-10}\,M_\odot$~yr$^{-1}$), which are close to the lower limit of the \textsc{Fried} grid (red solid line, $\dot{M}_\mathrm{wind}=10^{-10}\,M_\odot$~yr$^{-1}$), we show the median ratio of the two mass-loss rates for each region. $\sigma$ Orionis has a median $\dot{M}_\mathrm{wind}/\dot{M}_\mathrm{acc} =1.2$ (orange line) and the ONC core has $\dot{M}_\mathrm{wind}/\dot{M}_\mathrm{acc} =2.2$ (blue line). The black line represents $\dot{M}_\mathrm{wind} = \dot{M}_\mathrm{acc}$ (equilibrium between angular momentum transport and wind mass-loss).} }
         \vspace{-0.3cm} 
    \label{fig:mdotaccvwind}
\end{figure}

\begin{figure*}
    \includegraphics[width=0.84\textwidth]{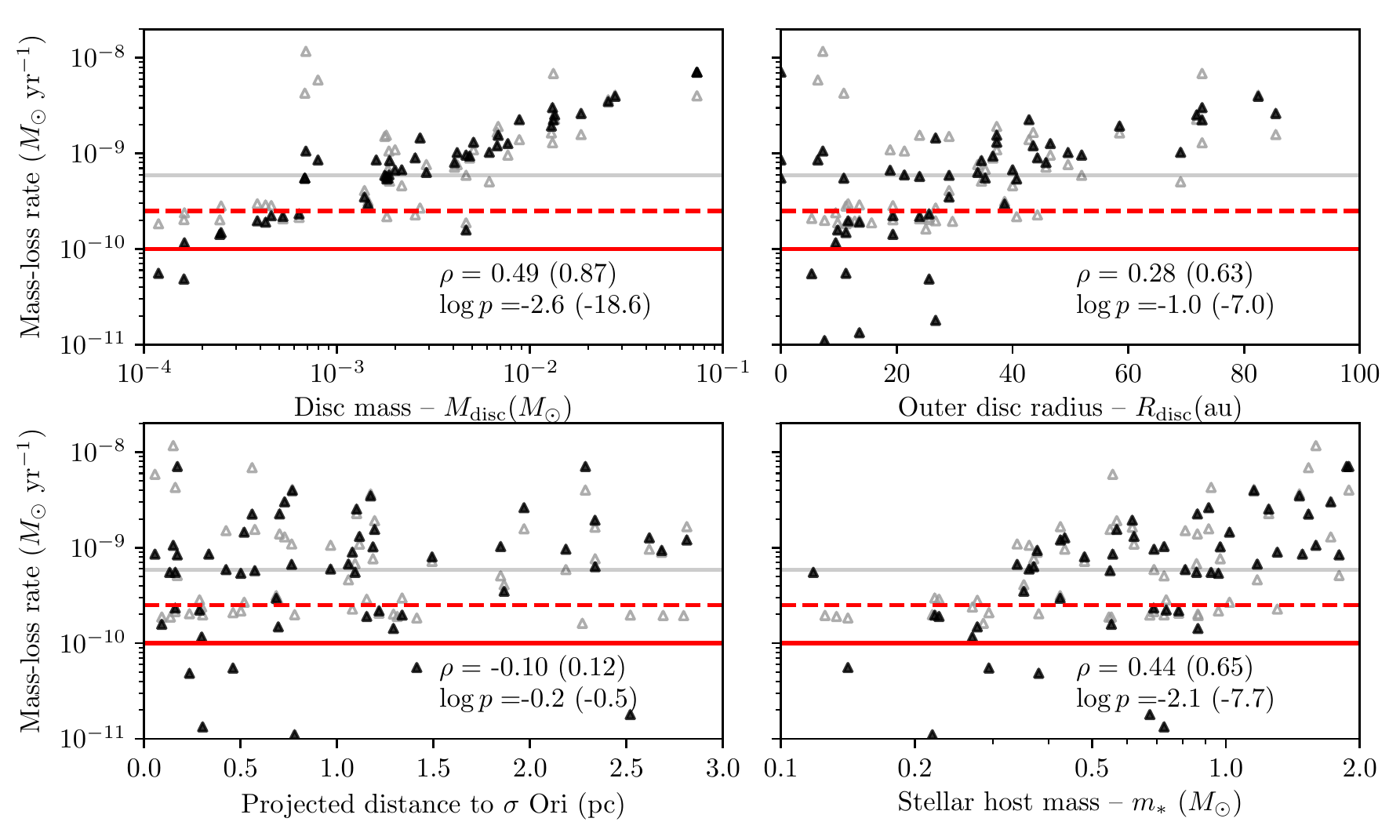}
    \vspace{-0.3cm}    
     \caption{Stellar accretion (solid) and photoevaporative (hollow) disc mass-loss rates as a function of disc mass, disc outer radius, projected distance from $\sigma$ Ori and stellar host mass. The median accretion rate is shown as faint horizontal black lines. The red horizontal line in each case marks the floor of the \textsc{Fried} grid, for which we assume $\dot{M}_\mathrm{wind}=0$ in our models, allowing the free viscous expansion of the outer radius. The Spearman's rank correlation coefficients, $\rho$, and corresponding (logarithmic) $p$-values are shown for the each property with $\dot{M}_\mathrm{wind}$, excluding data points for which $\dot{M}_\mathrm{wind}< 2.5 \times 10^{-10} \, M_\odot$~yr$^{-1}$ (below the red dashed line). In brackets we show the same values for \textit{all} $\dot{M}_\mathrm{acc}$. }
         \vspace{-0.4cm} 
     \label{fig:mlossrates}
   \end{figure*}

\subsection{Accretion and wind-driven mass-loss rates}
\label{sec:mdot}

Here we predict the distribution of photoevaporative mass-loss rates in $\sigma$ Orionis. This is intended as both a target selection guide for future observational studies, and a demonstration that a sufficient sample of directly measured mass-loss rates will provide valuable constraints of PPD physics. 

\vspace{-0.3cm}
\subsubsection{Comparing wind and accretion mass-loss rates}
\label{sec:mdotwindacc}

We are motivated to investigate how the physics of angular momentum transport, responsible for replenishing disc material in the wind-launching region in the outer part of the disc, influences the expected photoevaporative mass-loss rate. Our models assume viscous diffusion of the disc such that the mass flux outwards is comparable to the accretion rate onto the central star. While initially an extended disc in a region exposed to strong FUV field may undergo (rapid) outer radius depletion \citep{Win18b}, discs with truncated radii also experience lower mass-loss rates \citep{Haw18b}. Eventually, the wind driven mass-loss rate must therefore balance with any mass flux at the outer edge due to angular momentum transport. For a viscous disc, such an `equilibrium' state yields a wind driven mass-loss rate $\dot{M}_\mathrm{wind} \sim \dot{M}_\mathrm{acc}$, with some distribution around this value expected due to stellar dynamics (fluctuations in UV flux).

In Figure~\ref{fig:mdotaccvwind}, we show the distribution of $\dot{M}_\mathrm{acc}$ versus $\dot{M}_\mathrm{wind}$ for the disc population in our models of $\sigma$ Orionis, compared to those from similar models for the core of the younger Orion Nebula Cluster (ONC) by \citet{Win19b}. In the ONC, the young stellar population means that many PPDs are undergoing rapid truncation such that $\dot{M}_\mathrm{wind}/\dot{M}_\mathrm{acc} \gg 1$; these are the brightest proplyds with the highest mass-loss rates that have traditionally been targeted for observations \cite[see][]{Win19b}. However, the median ratio for all discs is $\dot{M}_\mathrm{wind}/\dot{M}_\mathrm{acc} = 2.2$, and the population at large is therefore undergoing an epoch of depletion of the outer radius, albeit less rapidly than the extreme proplyds with $\dot{M}_\mathrm{wind}\gtrsim 10^{-7}\,M_\odot$~yr$^{-1}$. For $\sigma$ Orionis, the ratio is distributed with median $\dot{M}_\mathrm{wind}/\dot{M}_\mathrm{acc} = 1.2$ with fewer extreme examples for which $\dot{M}_\mathrm{wind}/\dot{M}_\mathrm{acc}\gg 1$.  For the older photoevaporating disc population, the accretion rate should therefore be a strong indicator of the expected photoevaporation rate if viscous diffusion is operative. Conversely, if constraints on mass-loss rates indicate $\dot{M}_\mathrm{wind} \ll \dot{M}_\mathrm{acc}$ for many discs in the region, this would suggest that discs are not subject to viscous diffusion. Finding correlations between $\dot{M}_\mathrm{acc}$ and $\dot{M}_\mathrm{wind}$ in younger regions such as the ONC would also contrain angular momentum transport in PPDs, although $\dot{M}_\mathrm{wind}/\dot{M}_\mathrm{acc}>1$ is expected in this case, and care must be taken not to select only the brightest proplyds.

\vspace{-0.3cm}
\subsubsection{Variation with host and disc properties}
We are further interested in how $\dot{M}_\mathrm{wind}$ depends on stellar host and disc properties to motivate targets for future observational study. In Figure~\ref{fig:mlossrates} we show $\dot{M}_\mathrm{wind}$ and $\dot{M}_\mathrm{acc}$ from our models for disc mass, outer radius, projected separation from $\sigma$ Orionis and stellar host mass. We calculate the Spearman $R$-statistic for both mass-loss rates with respect to each property, only including PPDs with photoevaporative mass-loss rates of $\dot{M}_\mathrm{wind}>2.5\times 10^{-10}\, M_\odot$~yr$^{-1}$ in the case of correlations with $\dot{M}_\mathrm{wind}$.

Both $\dot{M}_\mathrm{wind}$ and $\dot{M}_\mathrm{acc}$ correlate with the total mass of PPDs and stellar host mass, since a greater surviving disc mass implies a greater outwards mass-flux in the disc. In addition, the models predict that disc outer radius should be correlated with $\dot{M}_\mathrm{acc}$ (and possibly with $\dot{M}_\mathrm{wind}$), since a greater outwards mass-flux is balanced with $\dot{M}_\mathrm{wind}$ at a greater radial disc extent.

Counterintuitively, neither $\dot{M}_\mathrm{wind}$ nor $\dot{M}_\mathrm{acc}$ correlates with proximity to $\sigma$ Ori. While FUV flux increases at small separations, this is balanced by the preferential depletion of close-in PPDs such that $\dot{M}_\mathrm{wind}$ is not strongly dependent on separation. In our models $\dot{M}_\mathrm{acc}$ increases with increasing $M_\mathrm{disc}$ and the maximum $M_\mathrm{disc}$ increases with separation from $\sigma$ Ori. A second order correlation may therefore be expected between $\dot{M}_\mathrm{acc}$ and the separation. We find no such correlation is detectable. This is because the relationship between $\dot{M}_\mathrm{acc}$ and ${M}_\mathrm{disc}$ is dominated by discs that lie at separations $\gtrsim 0.5$~pc (the majority), where there remains a broad distribution of ${M}_\mathrm{disc}$. Physically, this finding suggests that there is no strong gradient in the externally driven depletion of the inner disc. This does not mean external photoevaporation has a negligible overall influence on PPDs in the region.

\citet{Rig11} used $U$-band photometry to estimate accretion rates in $\sigma$ Orionis. Photometrically derived accretion rates are less certain than those obtained from spectroscopic measurements, and many of the constraints on $\dot{M}_\mathrm{acc}$ are upper or lower limits. Nonetheless, the authors find that $\dot{M}_\mathrm{acc}$ in the region is positively correlated with stellar host mass and uncorrelated with distance from $\sigma$ Ori, in agreement with the results of our model. However, the correlation between $M_\mathrm{disc}$ and $\dot{M}_\mathrm{acc}$ cannot be conclusively demonstrated due to the lack of overlap between the detected sub-mm flux sample and stars with constraints on accretion rates \citep[see discussion in][]{Ans17}. The observed median accretion rate in $\sigma$ Orionis is $\sim 3 \times 10^{-10}\, M_\odot$~yr$^{-1}$, a factor $\sim 2$ lower than those obtained in our model. However, since many of the measured $\dot{M}_\mathrm{acc}$ are limits, this value should be interpreted with caution; this may be improved by future spectroscopically derived accretion rates \citep[e.g. with \textit{X-shooter} --][]{Man17}. In addition, the \citet{Rig11} sample contains fewer high mass host stars than our model ($\gtrsim 1 \, M_\odot$ -- i.e. those with highest predicted accretion rates). This may be due to internal dispersal mechanisms acting more rapidly for PPDs around such stars \citep[e.g.][]{Rib15}.

We conclude that future observations aimed at constraining $\dot{M}_\mathrm{wind}$ for a depleted population of PPDs should prioritise the most massive discs around the most massive stars with the highest stellar accretion rates that are close to an ionising source. 

\vspace{-0.3cm}
\subsection{Caveats}
\label{sec:caveats}

Our simplified model has made a number of assumptions that we mention here briefly. Firstly, when drawing comparisons between the disc masses in our model and those inferred from {ALMA} observations, we have simply assumed a dust-to-gas ratio of $10^{-2}$. This ratio may be influenced by the preferential evaporation of gas and small dust grains, and numerous considerations for dust evolution may lead to properties that are distinct from the gas component \citep[see discussion by][]{Sel20}. In addition, we have ignored internal disc dispersal mechanisms, such as photoevaporative/magnetic winds, that likely play a role for an intermediate age disc population \citep{Erc17}. We have also fixed the viscous $\alpha=10^{-3}$; increasing it would decrease the disc masses at the present time such that it is somewhat degenerate with the assumed age of the region \citep[see e.g.][for studies varying $\alpha$]{Ros17, Win19b}. Finally, we have not included any prescription for interstellar extinction that may shield discs from UV irradiation, reducing mass-loss at early times \citep{Ali19}. None of these considerations affect the prediction that a sufficiently depleted disc population must exhibit a balance between externally driven winds and mass flux due to angular momentum transport. 

\vspace{-0.3cm}
\section{Conclusions}

We have presented a model for the externally driven photoevaporation of PPDs in the intermediate age star-forming region, $\sigma$ Orionis. The model reproduces the observed disc mass distribution, suggesting that this is a viable mechanism for the depletion of circumstellar material in the region. Since externally driven disc winds are launched from the outer edge of the disc, photoevaporative mass-loss rates are eventually moderated by the outward viscous mass flux in these outer regions. This means a region of depleted PPDs can be used to put constraints on angular momentum transport processes. In particular, if viscous diffusion operates to redistribute angular momentum then $\dot{M}_\mathrm{wind} \sim \dot{M}_\mathrm{acc}$ across the population, while younger regions such as the ONC should exhibit systematically larger $\dot{M}_\mathrm{wind}$ than $\dot{M}_\mathrm{acc}$ (indicating ongoing disc truncation).  If instead angular momentum is extracted from the disc (e.g. by magnetohydrodynamic winds) then $\dot{M}_\mathrm{wind} \ll \dot{M}_\mathrm{acc}$ is expected for many discs in regions such as $\sigma$~Orionis. If a sufficient sample of mass-loss rates can be directly measured using disc wind tracers \citep[e.g. atomic carbon --][]{Haw20}, then in conjunction with accretion rates this can be used to constrain angular momentum transport in the outer regions of a PPD.

\vspace{-0.3cm}
\section*{Acknowledgements}
{We thank the reviewer, Cathie Clarke, for a useful report which helped to significantly improve this manuscript.} AJW acknowledges funding from an Alexander von Humboldt Stiftung Postdoctoral Research Fellowship and the European Research Council (ERC) under the European Union's Horizon 2020 research and innovation programme (grant agreement no.\ 681601). MA acknowledges support from NASA grant NNH18ZDA001N/EW. TJH is funded by a Royal Society Dorothy Hodgkin Fellowship. JMDK acknowledges funding from the Deustsche Forschungsgemeinschaft (DFG, German Research Foundation), grant numbers KR4801/1-1 and KR8401/2-1 and from the European Research Council (ERC) under the European Union's Horizon 2020 research and innovation programme (grant agreement no.\ 714907).

\vspace{-0.5cm}



\bibliographystyle{mnras}
\bibliography{truncation} 
\vspace{-0.2cm}



\bsp	
\label{lastpage}
\end{document}